\documentclass[cits]{PoS}
\usepackage{wrapfig}

\title{Complete nucleosynthesis calculations for low$-$mass stars from NuGrid}

\ShortTitle{Nucleosynthesis for low$-$mass stars}

\author{
\speaker{M. Pignatari}$^{abc}$,
F. Herwig$^{abd}$,
M. E. Bennett$^{ad}$,
S. Diehl$^{aef}$,
C. L. Fryer$^{af}$,
R. Hirschi$^{abg}$,
A. Hungerford$^{af}$,
G. Magkotsios$^{ach}$,
G. Rockefeller$^{af}$,
F. X. Timmes$^{ah}$,
and P. Young$^{ah}$\\
\llap{$^a$}The NuGrid Collaboration\\
\llap{$^b$}Astrophysics Group, Keele University, ST5 5BG, UK\\
\llap{$^c$} Joint Institute for Nuclear Astrophysics, University of Notre Dame, IN, 46556, USA\\
\llap{$^d$}Dept. of Physics \& Astronomy, Victoria, BC, V8W 3P6, Canada\\
\llap{$^e$}Theoretical Astrophysics Group (T-6), Los Alamos National Laboratory, Los Alamos, NM, 87544, USA\\
\llap{$^f$}Computational Methods (CCS-2), Los Alamos National Laboratory, Los Alamos, NM, 87544, USA\\
\llap{$^g$}IPMU, University of Tokyo, Kashiwa, Chiba 277-8582, Japan\\
\llap{$^h$}School of Earth and Space Exploration, Arizona State University, Tempe, AZ 85287, USA\\
E-mail:\email{marco@astro.keele.ac.uk}
}
\abstract{Many nucleosynthesis and mixing processes of low$-$mass stars as they evolve from
the Main Sequence to the thermal-pulse Asymptotic Giant Branch phase 
(TP$-$AGB) are well understood (although of course important physics components, 
e.g. rotation, magnetic fields, gravity wave mixing, remain poorly known). 
Nevertheless, in the last years presolar grain measurements with high 
resolution have presented new puzzling problems and strong constraints 
on nucleosynthesis processes in stars. 
The goal of the NuGrid collaboration is to present uniform yields for a large
range of masses and metallicities, including low$-$mass stars and massive stars
and their explosions. Here we present the first calculations of 
stellar evolution and high$-$resolution, post$-$processing simulations of 
an AGB star with an initial mass
of 2 M$_{\odot}$ and solar$-$like metallicity (Z=0.01), 
based on the post$-$processing code $PPN$.
In particular, we analyze the formation and evolution of the radiative
$^{13}$C$-$pocket between the 17th TP and the 18th TP. The $s$-process 
nucleosynthesis profile of a sample of heavy isotopes is also discussed, before
the next convective TP occurrence.}

\FullConference{10th Symposium on Nuclei in the Cosmos\\
		 July 27 - August 1 2008\\
		 Mackinac Island, Michigan, USA}

\begin{document}

\section{Introduction}

When He$-$burning is exhausted in the core, low mass stars 
(1.5 $-$ 3 M$_{\odot}$) evolve along the AGB. 
Late on the AGB, recurrent thermal instabilities called Thermal Pulses 
(TP-AGB phase, 
%Schwarzschild \& H\"arm 1965, 
\cite{schwarzschild:65}) affect shell He$-$burning history. 
After TPs (time scale in the order of few hundreds years), the third 
dredge$-$up events (TDU) mix He shell material in the envelope, and fresh 
protons down in the He$-$intershell. 
A $^{13}$C$-$pocket is formed in the radiative He intershell phase, 
where the $^{13}$C($\alpha$,n)$^{16}$O neutron source becomes 
efficient activating slow neutron capture process 
(s$-$process, \cite{burbidge:57}). 
A marginal contribution is also given by the partial activation of the 
$^{22}$Ne($\alpha$,n)$^{25}$Mg at the bottom of the He intershell
(e.g., \cite{gallino:98}).\\
As a result of the TDU enriching the AGB envelope with $s$-process rich material,
AGB stars provide most of the s elements beyond Sr observed in the Solar System.
In particular, the "main component" between Sr and Pb is produced by 
solar$-$like AGB stars, while the "strong component" explaining half of the 
solar $^{208}$Pb is produced by low metallicity AGB stars 
(\cite{arlandini:99}, and references therein). 
Carbon is also dredged-up with s elements in the envelope, 
and eventually the AGB star may become a C-rich star (C(N) star), 
meaning that carbon is more abundant than oxygen in the envelope.\\ 
Spectroscopic observations and composition measurements in presolar 
grains formed in AGB stars confirm this scenario 
(\cite{busso:01},\cite{lugaro:03b} ,respectively), 
and provide important insight to study and understand those stars in more 
details. In particular, presolar grains carry the isotopic and chemical 
signature of their parent stars (e.g., \cite{zinner:98},\cite{lugaro:03b},
\cite{barzyk:06}), 
providing a powerful tool to test and constrain stellar models and nuclear 
physics inputs. 
The $NuGrid$ project (see also Herwig et al. in this volume) 
has the goal to generate uniform yields for a large range of masses 
and metallicities also for low$-$mass stars, and to constrain them with 
observations.
In this proceeding we present the first calculations of stellar evolution and 
high-resolution, post-processing simulations of a 2 M$_{\odot}$ Z = 0.01 AGB 
star, based on the post$-$processing code $PPN$.

\section{Post$-$processing calculations}

The main input parameters for the post-processing calculations are given by
a 2 M$_{\odot}$ and Z = 0.01 star (EVOL Code, e.g., \cite{herwig:06}).
The stellar model has been calculated assuming an overshoot parameter 
$f$ = 0.128 at the bottom of the envelope and $f$ = 0.008 for all the other
convective boundaries.
The $f$ applied at the base of the convective TP
in the He shell ($f$ = 0.008) has been constrained to explain the He/C/O ratio 
observed in H deficient post-AGB stars of type PG1159 and in WC central stars 
of planetary nebulae (\cite{werner:06}). 
The higher overshoot parameter at the bottom of the envelope is 
calibrated to reproduce  the mass of the $^{13}$C-pocket needed to reproduce 
the observed overabundance of $s$-process elements \cite{lugaro:03a}. 
The post-processing code $PPN$ is described in Herwig et al. (this vol.) 
and includes dynamically all species from H to Bi.
Concerning the simulations shown in this proceeding, 
from the physics package in $PPN$ we selected 
\cite{angulo:99} (NACRE compilation) for the main charged particle reactions,
\cite{dillmann:06} (Kadonis compilation) for neutron capture reactions 
and \cite{rauscher:00} for unstable isotopes not
included in the Kadonis network.
We selected \cite{oda:94} and \cite{fuller:85} (or terrestrial rates if not 
available in the previous references) for stellar $\beta$-decay rates of light 
unstable isotopes, and \cite{goriely:99} (or terrestrial rates if not 
available) for $\beta$-decay rates of heavy unstable isotopes 
(see also Herwig et al. in this volume for more details about the 
physics package).\\
\begin{figure}[h!]
 \includegraphics[angle=-90,width=0.5\textwidth]{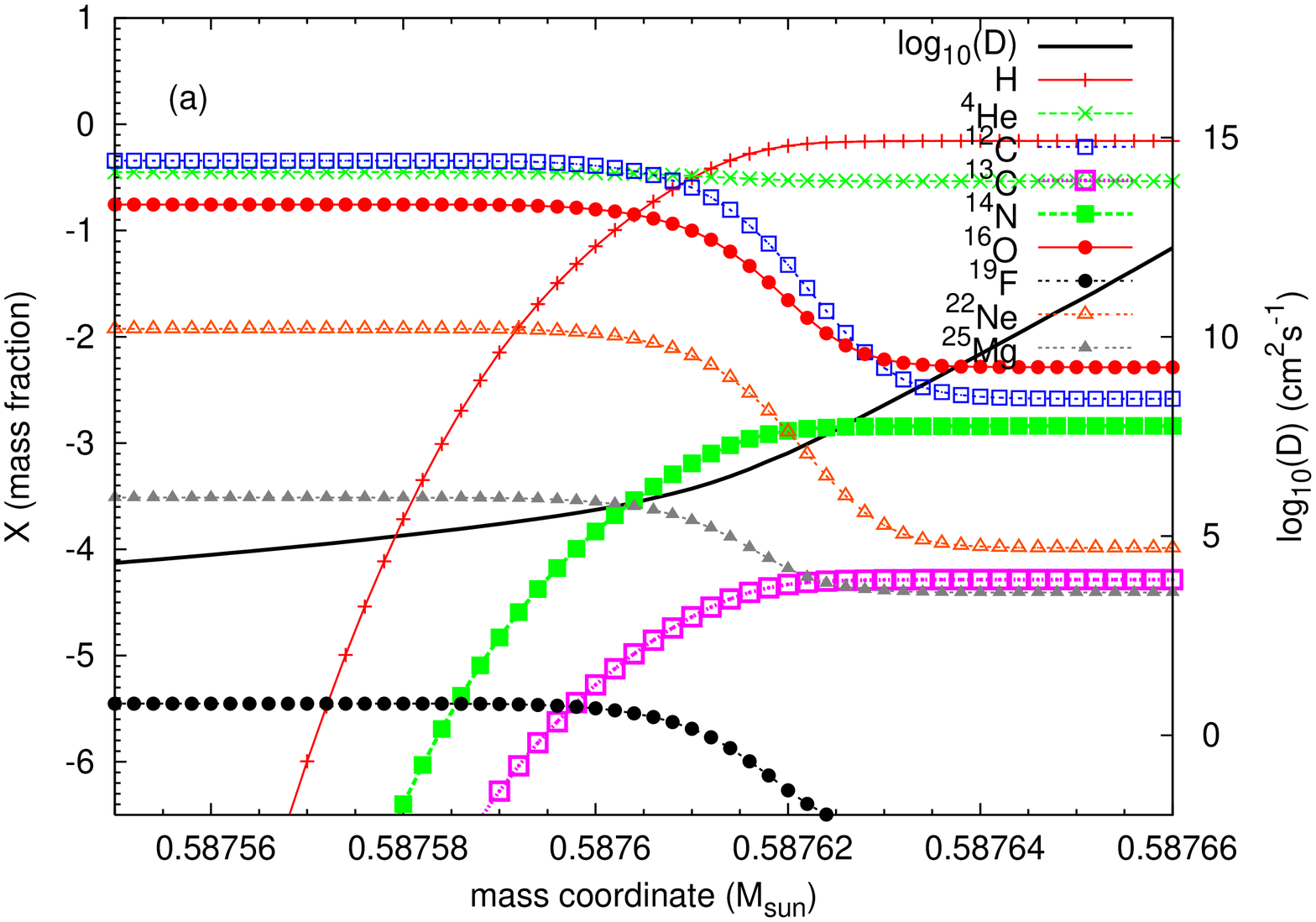}
 \includegraphics[angle=-90,width=0.5\textwidth]{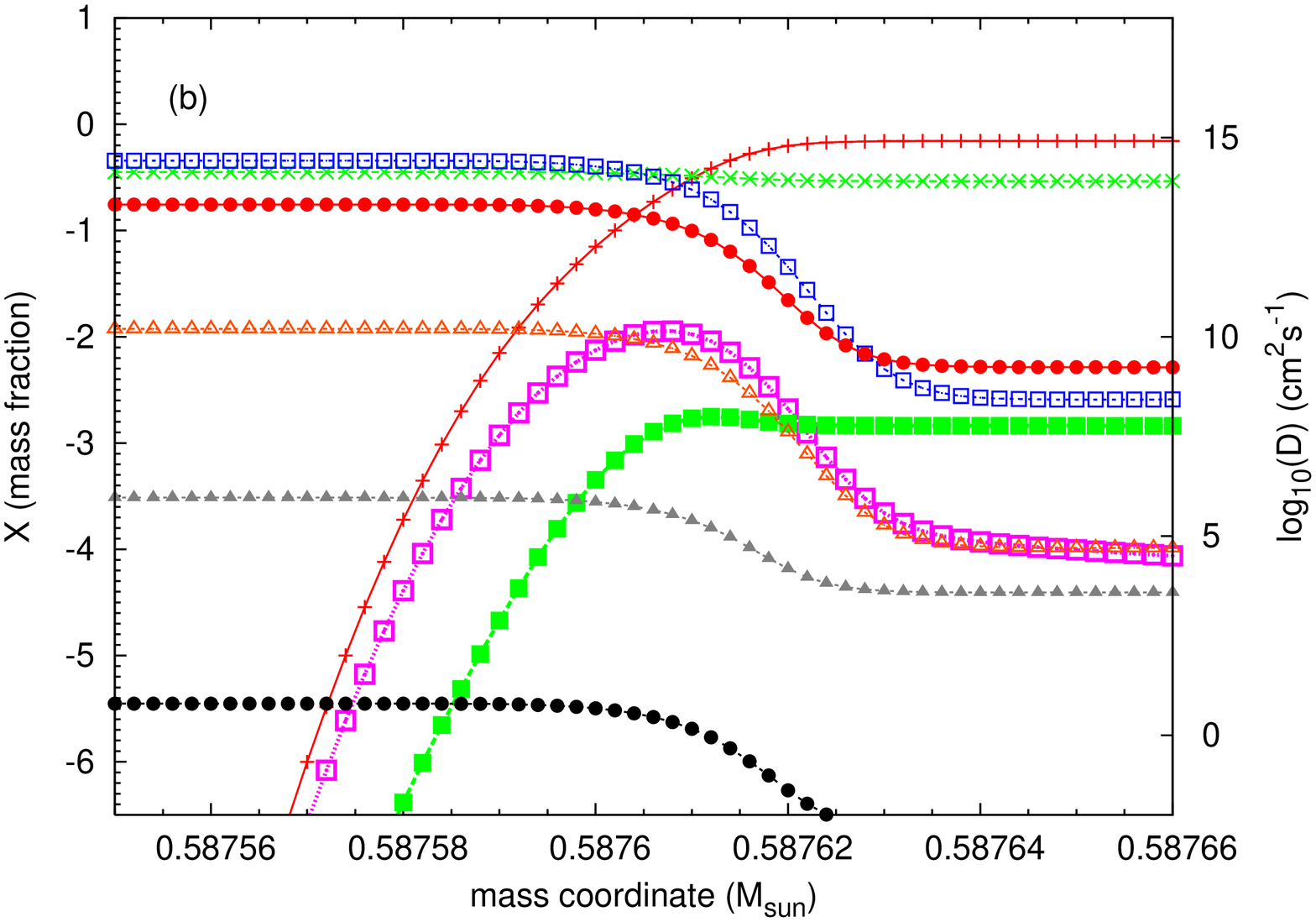}
 \includegraphics[angle=-90,width=0.5\textwidth]{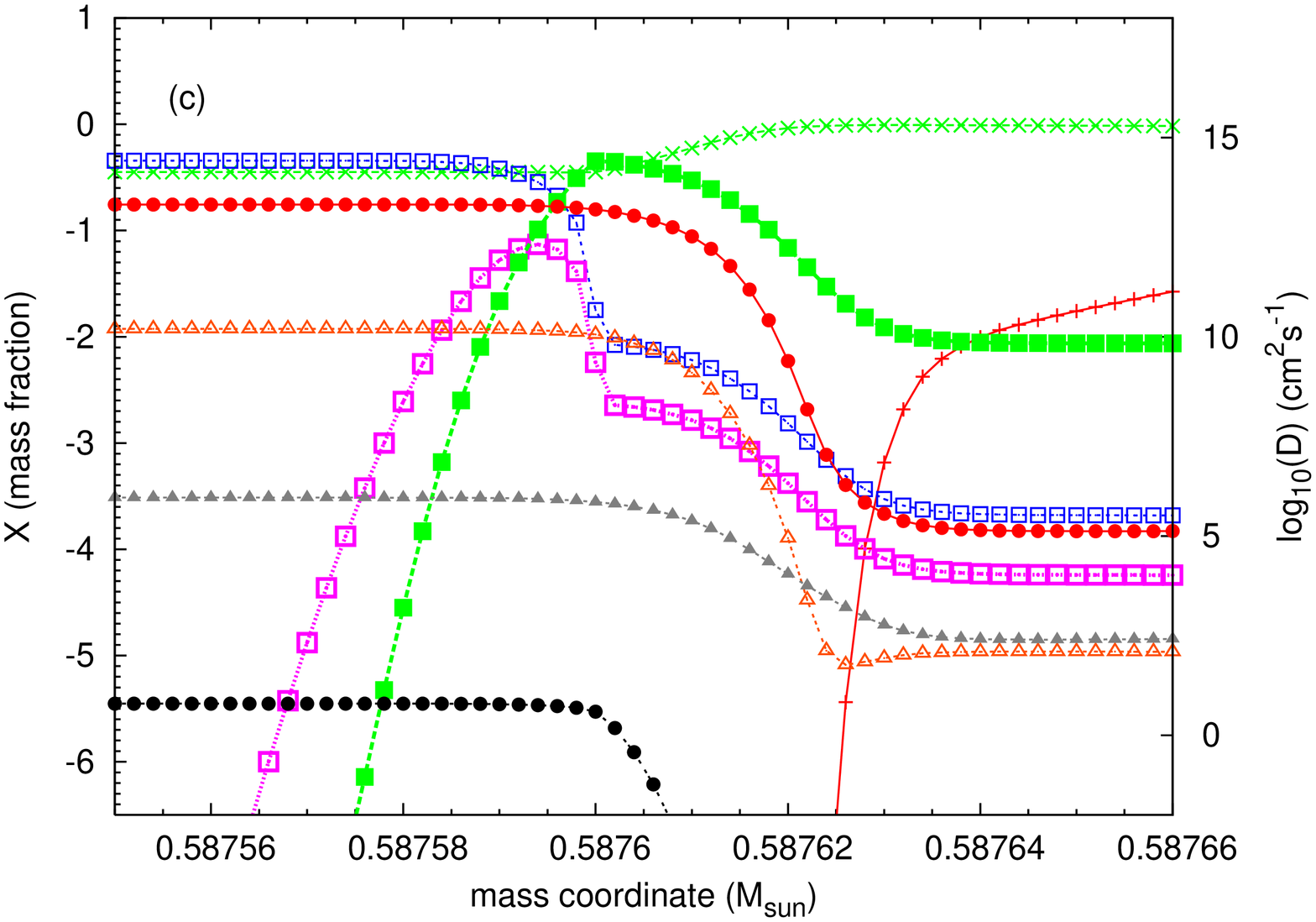}
 \includegraphics[angle=-90,width=0.5\textwidth]{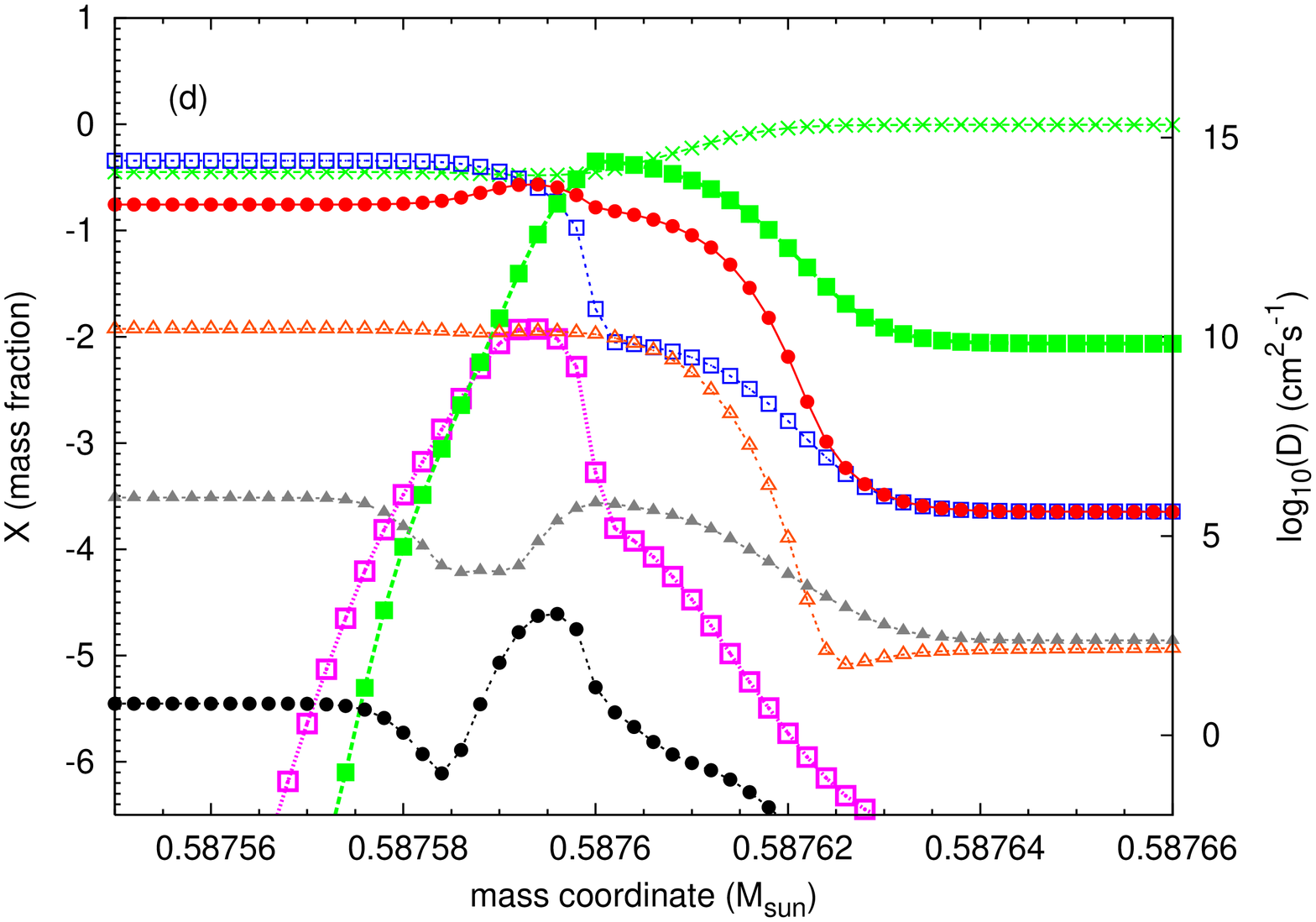}
 \caption{Panel (a,b,c,d): Formation and evolution of the $^{13}$C-pocket 
 after the 17th TP for the 2 M$_{\odot}$ Z = 0.01 star. 
 The profile for the mixing coefficient D (it is different from zero only in
 Panel (a)) and for a sample of light isotopes is provided. }
 \label{fig:C13pocket-evol}
\end{figure}
\begin{wrapfigure}{p}{0.5\textwidth}
\centering
 \includegraphics[angle=-90,width=0.5\textwidth]{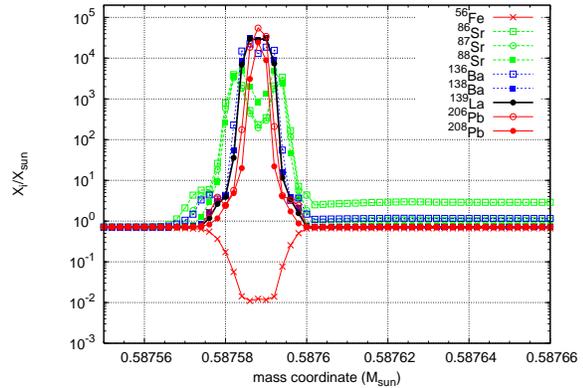}
 \caption{The overabundances of a sample of heavy isotopes has been reported
 (among them the s-only isotopes $^{86,87}$Sr, $^{136}$Ba and $^{206}$Pb),
 at the same time-step of Fig. 1, Panel (d).}
 \label{fig:abu}
\end{wrapfigure}
In Fig. \ref{fig:C13pocket-evol} we show the
formation and evolution of the $^{13}$C-pocket between the 
17th TP and the 18th TP (interpulse phase of about 70000 yr).
In Panel (a) the TDU mixes down in the He intershell protons and 
envelope material, before the re-activation of the H shell. 
In Panel (b) the abundant $^{12}$C is efficiently capturing protons 
producing $^{13}$C in radiative conditions via the nucleosynthesis channel 
$^{12}$C(p,$\gamma$)$^{13}$N($\beta^+$)$^{13}$C (e.g., \cite{gallino:98}). 
At the $^{13}$C abundance peak, 
also $^{14}$N starts to be produced by the proton capture channel 
$^{13}$C(p,$\gamma$)$^{14}$N. 
In Panel (c) the $^{13}$C-pocket final shape is shown, since
protons are fully consumed. The pocket size is 
1$-$2$\times$10$^{-5}$ M$_{\odot}$. Moving outward, a prominent $^{14}$N-pocket is
formed just after the $^{13}$C-pocket, as expected.
After about 40000 years, the temperature in the $^{13}$C-pocket is high enough to
efficiently activate the $^{13}$C($\alpha$,$n$)$^{16}$O reaction, producing
neutrons for the $s$ process.
In Panel (d) we show the $^{13}$C-pocket region once $^{13}$C has been burnt and
the $s$ process is not anymore efficient.
$^{25}$Mg is the main neutron poison in the $^{13}$C-pocket, and it has been
partially depleted by the neutron flux. 
In the $^{14}$N-pocket, instead, as it is well known 
$^{14}$N($n$,$p$)$^{14}$C is the main neutron poison and the neutron 
capture efficiency of $^{25}$Mg (and of the main $s$-process seed $^{56}$Fe) 
quickly decreases with increasing the $^{14}$N abundance.
The final $^{19}$F abundance profile basically follows the $^{13}$C profile.
If $^{13}$C is more abundant than $^{14}$N, then $^{19}$F is depleted by neutron
capture and by $\alpha$ capture.
In case $^{14}$N is more abundant than $^{13}$C, $^{19}$F can be produced
starting from $^{14}$N \cite{jorissen:92}. \\
Finally, in Fig. \ref{fig:abu}, we report the final 
overabundance profile in the $^{13}$C-pocket region for a sample of isotopes
at the Sr neutron magic peak ($^{86,87,88}$Sr), 
at the Ba neutron magic peak ($^{136,138}$Ba, $^{139}$La) and at 
the Pb neutron magic peak ($^{206,208}$Ba).
The $ls$ peak species (e.g. Sr, \cite{busso:01}) show a maximum of 
overproduction of about 5$\times$10$^3$, while the $hs$ peak (e.g. Ba) 
and the Pb peak show an overproduction of about few 10$^4$.
Sr isotopes show a double peak in coincidence of the 
$^{56}$Fe depletion tails. At 0.58758 M$_{\odot}$ the 
$^{13}$C abundance is rapidly decreasing and as
a consequence a lower amount of neutrons are produced. 
On the other hand, at 0.587595 M$_{\odot}$ the poisoning effect of 
$^{14}$N is increasing, until the $^{56}$Fe neutron capture efficiency is
negligible. The Ba peak and the Pb peak are more produced in the center of the  
$^{13}$C-pocket, where lighter Sr peak elements are feeding $s$ nucleosynthesis
of heavier elements. 
 The next convective TP will mix the $s$-process rich pocket in all the He
intershell, which will be partially dredged up in the envelope
by the next TDU event.\\ 

The analysis presented Fig. \ref{fig:C13pocket-evol} and in Fig. \ref{fig:abu}
shows part of the capabilities of the 
$PPN$ post-processing code applied to AGB nucleosynthesis calculations.
At present, we may calculate the abundances from H to Bi 
(including isotopic ratios) at any position and at any time in 
a complete stellar track.
Furthermore, in the nuclear network every reaction rate may be 
automatically chosen between different nuclear sources, 
or a multiplication factor can be applied or the reaction may be 
not considered. This opens up possibilities to systematically take into account 
the effect of nuclear reaction rate uncertainties in our nucleosynthesis
calculations.

\acknowledgments 

M.P. acknowledges support through NSF grants PHY
02-16783 (JINA). M.P. and F.H. were supported by a Marie Curie 
International Reintegration Grant MIRG-CT-2006-046520 within the European FP6.

\bibliographystyle{amsplain}
\bibliography{references}

\end{document}